\documentclass[twocolumn]{revtex4}
\usepackage{graphicx}
\usepackage{dcolumn}
\usepackage{bm}
\usepackage{color}
\usepackage{eso-pic}
\usepackage{amsmath}
\newcommand{\YBCO}{$\mbox{YBa}_2$$\mbox{Cu}_3$$\mbox{O}_7${$\mbox{ } $}}

\begin{document}

\title{Josephson mixers for terahertz detection}

\author{M. Malnou$^1$, C. Feuillet-Palma$^1$, C. Ulysse$^2$, G. Faini$^2$, P. Febvre$^3$, M. Sirena$^4$,L. Olanier$^1$, J. Lesueur$^1$, N. Bergeal$^1$}

\affiliation{$^1$Laboratoire de Physique et d'Etude des Mat\'eriaux - UMR8213-CNRS-ESPCI ParisTech-UPMC, 10 Rue Vauquelin - 75005 Paris, France.}
\affiliation{$^2$Laboratoire de Photonique et de Nanostructures LPN-CNRS, Route de Nozay, 91460 Marcoussis, France.}
\affiliation{$^3$IMEP-LAHC - UMR 5130 CNRS, Universit\'e de Savoie, 73376 Le Bourget du Lac cedex, France.}
\affiliation{$^4$Centro At\'omico Bariloche, Instituto Balseiro Ð CNEA and Univ. Nac. de Cuyo, Av. Bustillo 9500, 8400 Bariloche, Rio Negro Ð Argentina.}
\date{\today}

\begin{abstract}
 We report on an experimental and theoretical study of the high-frequency mixing properties of ion-irradiated \YBCO Josephson junctions embedded in THz antennas. We investigated the influence of the local oscillator power and frequency on the device performances.  The experimental data are compared with theoretical predictions of the general three-port model for mixers, in which the junction is described by the resistively shunted junction model. A good agreement is obtained for the conversion efficiency in different frequency ranges, spanning above and below the characteristic frequencies $f_c$ of the junctions.
 \end{abstract}

\maketitle
 
The THz region of the electromagnetic spectrum, which covers the range from 0.3 to 10 THz is a frontier area for research in many fields, including physics, astronomy, chemistry, material science and biology. However, so far, this range is hardly exploitable because of the limited number of suitable sources and detectors \cite{ferguson,tonouchi}. Indeed, THz frequencies lie between the frequency range of electronics and photonics where the existing technologies cannot be simply extended.  An important challenge for practical applications is heterodyne detection,  that is needed to translate the THz frequency window of interest to lower frequencies, for which semiconductor electronics can process the signals. This detection technique has the advantage to combine high sensitivity with high frequency resolution. It involves detecting a signal at frequency $f_s$ by non-linear mixing with a continuous wave reference signal produced by a local oscillator (LO) at frequency $f_{\mathrm{LO}}$. The output signal has a low frequency component at the intermediate frequency  (IF) $f_{\mathrm{IF}}=|f_{\mathrm{LO}}-f_s|$ which contains the information carried by the original signal.  Metal-semiconductor Schottky diodes are often employed as mixing elements in heterodyne receivers, offering ease of use and a wide coverage of the THz frequency region \cite{crowe,yasui}. Their main drawbacks are  the limited sensitivity and the need for a high LO power. Superconducting hot electron bolometers (HEB) made of Niobium or Niobium Nitride (NbN) are able to operate at low noise with very good frequency coverage but require cooling to 4K \cite{gershenzon, zmuidzinas,zhang}. So far, the most sensitive frequency-mixing elements are the low temperature superconductor-insulator-superconductor (SIS ) Niobium tunnel junctions \cite{graauw, mears}. However, these junctions are intrinsically limited in frequency by the gap energy of Nb ($\sim$800 GHz ) and operate only at low temperature (4 K). More recently, NbN SIS junctions with higher gap energy have been developed to extend the frequency range. \\
  \begin{figure}[t]
\includegraphics[width=8.8cm]{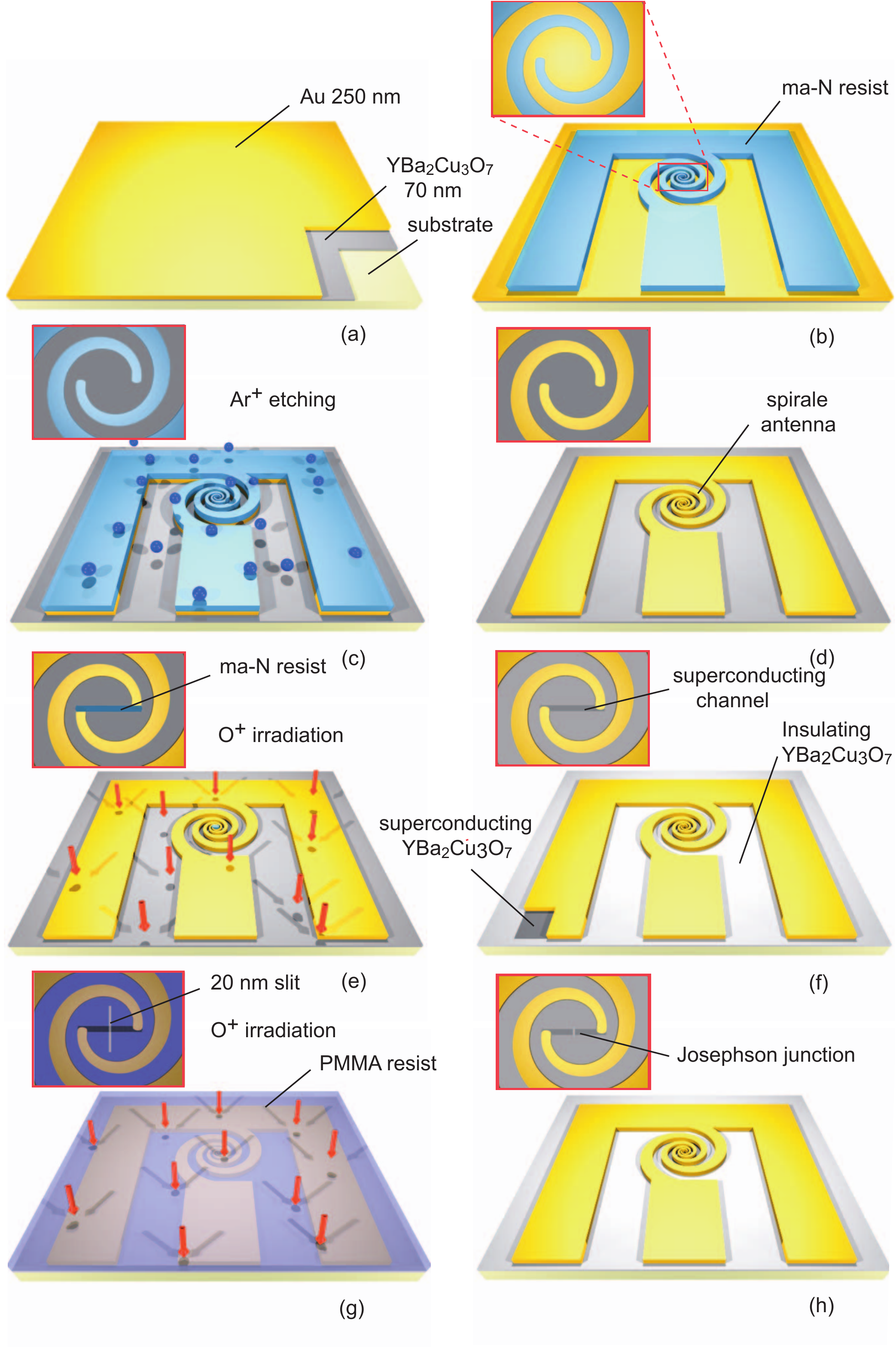}
    \caption{ Illustration of the fabrication process steps: (a) 70-nm thick \YBCO film grown on sapphire covered by an in situ 250-nm gold layer; (b) spiral antenna in the CPW transmission line defined in an ma-N negative e-beam resist; (c) 500-eV Ar ion-beam-etching of the gold layer (d) gold antenna in the CPW transmission line on \YBCO; (e) high-dose 70-keV oxygen ion irradiation to create insulating regions in exposed \YBCO. A 2-$\mu$m wide channel in the center of the antenna is protected by a  ma-N resist mask; (f) patterned superconducting and insulating \YBCO regions; (g) low-dose 110 keV oxygen ion irradiation of the Josephson junction patterned as a 20-nm-wide slit in PMMA photoresist; and (h) device after resist cleaning.}
 \end{figure}
	An alternative to these devices consist in using high-temperature superconducting   (HTS) receivers. In addition to the obvious advantage of a much higher operating temperature, their higher energy gap results in a cut-off frequency of several THz.  Hence, it is important to develop HTS devices and related heterodyne mixer technology for applications in the THz range. However, SIS tunnel junction technology is not available with these materials and it is not possible to directly adapt the low-T$_c$ superconductor technology.  In this context, the interest for non-tunnel Josephson junction mixers has been renewed because they can be fabricated by various methods with high-T$_c$ materials. Unlike SIS mixers, whose operation is based on the quasiparticle non-linearity near the gap energy, Josephson mixers use the non-linearity of the Cooper pair current. First realizations of mixers with high-T$_c$ superconductors which were mainly based on grain-boundary or ramp edge junctions \cite{chen,scherbel,harnack,tarasov}, produced promising results but the development was slowed down by the difficulty to build a junction technology sufficiently reliable to fabricate complex devices. In recent years, a new approach based on ion irradiation has been developed to fabricate Josephson junctions with high temperature superconductors. This method has been used to produce reproducible junctions \cite{kahlmann,bergeal}, SQUIDs \cite{bergealsq} and large-scale integrated Josephson circuits \cite{cybart1, cybart}. Here, we present an analysis of the high-frequency mixing properties of Josephson junctions made by ion irradiation. The experimental data are compared with theoretical predictions form the general three-port model. \\

\textbf{ I. Fabrication of the Josephson mixer.}\\

HTS thin films are structured at the nanometer scale by combining e-beam lithography with ion irradiation. This technique relies on the extreme sensitivity of HTS to defects, owing to the d-wave symmetry of their order parameter. Disorder induced in the material  by irradiation reduces the superconducting transition temperature and increases the resistivity because of enhanced scattering. Beyond a critical defect density, a superconductor-to-insulator transition takes place, a phenomenon that can be used to selectively insulate some regions of a superconducting film \cite{bergealjap}. Figure 1 describes the different steps of the fabrication process. Starting from a commercial 70-nm-thick \YBCO film ($T_c$ = 86 K) \cite{ceraco} grown on sapphire covered by an in-situ 250-nm gold layer (Fig.1a), a three steps fabrication process is performed. The spiral antenna embedded in a 50$\Omega$ co-planar waweguide (CPW) transmission line is first defined in the gold layer through a ma-N e-beam resist patterning followed by a 500-eV Ar$^+$ Ion Beam Etching (IBE) (Fig.1b,c,d). Then a 2-$\mu m$ wide channel  located at the center of the antenna is patterned in a ma-N e-beam resist, followed by a 70-keV oxygen ion irradiation at a dose of 2$ \times$10$^{15}$ at/cm$^2$ (Fig.1e). This process ensures that the regions of the film which are not protected either by the resist or by the gold layer become deeply insulating (Fig. 1f). No HTS material is removed and the superconducting parts, including the antenna, the CPW line and the 2-$\mu m$ wide channel, remain embedded in the insulating film preventing degradation. Finally, the junction is defined at the center of the superconducting channel by irradiating through a 20-nm  wide slit patterned in a PMMA resist with 100 keV oxygen ions (Fig.1g,h). A fluence of 3$ \times$10$^{13}$ at/cm$^2$ is used to lower the $T_c$ in the region underneath the slit. The parameters of the junctions such as the normal resistance $R_n$, the critical current  $I_c$ and the operating temperature can be engineered simply by modifying the width of the slit, the fluence of irradiation and the ion energy.\\ 
\indent Josephson behaviour, in particular the Fraunhofer pattern of the critical current under magnetic field and Shapiro steps under microwave irradiation  has been reported  previously in this type of junctions \cite{lesueur}. One main advantage of this technique is that the process is by nature highly scalable, with no design constraint. It is therefore particularly suitable in creating THz devices that include Josephson junctions embedded in their circuitry  \cite{malnou}. In this device, the self-complementary spiral wideband antenna  [80GHz-6THz] has an impedance $Z=\frac{Z_0}{2}\sqrt{\frac{2}{\epsilon_r+1}}\approx$ 80 $\Omega$ where $Z_0\approx$ 377$\Omega$ is the vacuum impedance and $\epsilon_r\approx$ 10 is the dielectric constant of the sapphire substrate. The 50-$\Omega$ CPW transmission line is directly connected to the junction to readout the intermediate frequency signal.\\

   \begin{figure}[h]
\includegraphics[width=8.5cm]{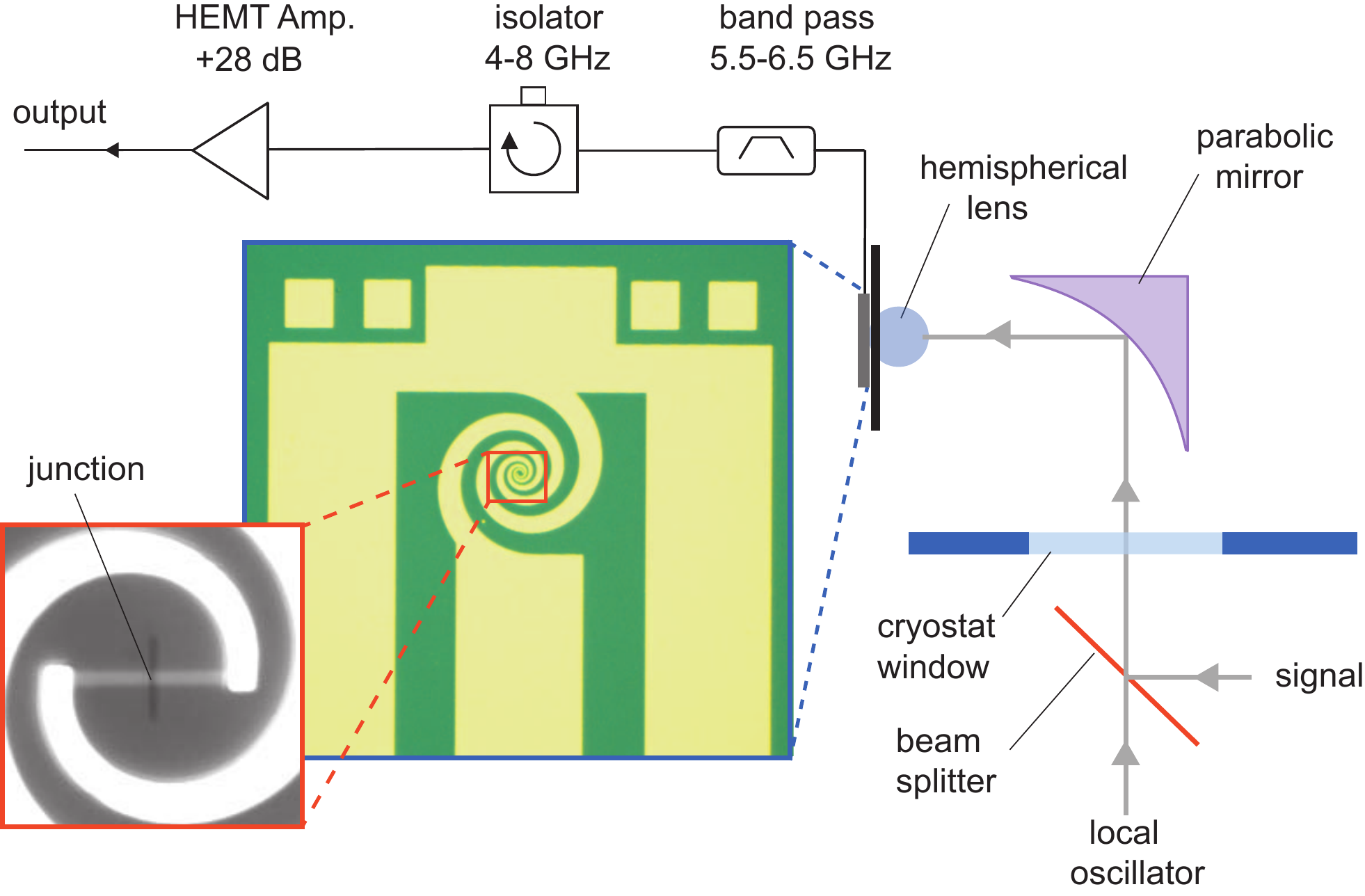}
    \caption{Josephson mixer in its quasi-optical and microwave set-up. The junction is embedded in a wide-band spiral antenna and is connected to a CPW transmission line.}
 \end{figure}

\textbf{ II. Experimental set-up}\\

The experimental set-up is shown in Fig. 2. The back side of the sapphire substrate is placed in contact with a silicon hyper-hemispheric lens located at the focal point of a parabolic mirror exposed to external signals through the window of the cryostat. The junction is connected to contact pads for dc biasing and to a microwave transmission line. A cryogenic HEMT amplifier operating in the 4-8GHz band amplifies the output signal at the intermediate frequency before further amplification at room temperature. An isolator is placed in the chain to  minimize the back-action of the amplifier on the Josephson mixer. The local oscillator is combined with the signal through a  beam splitter. Mixing experiments were performed in five different frequency ranges centered on 20, 70, 140, 280 and 420 GHz. At 20 GHz, signals are provided by  microwave generators whereas for the higher frequencies, signals are provided by Gunn diodes emitting at 70GHz coupled to a set of  frequency doublers and triplers. \\

\textbf{III. dc and ac response of the junction}\\

\indent 
The  resistance of the junction as a function of temperature measured at very low current bias, reveals the existence of two characteristic temperatures in our device, namely $T_{c}$ and $T_J$ (fig. 3a). The highest transition at $T_{c}$  = 84 K refers to the superconducting transition of the non-irradiated regions of sample, which corresponds to the transition temperature of the unprocessed \YBCO film  \cite{bergeal}.  The second transition at the lower temperature $T_{J}$ = 66 K corresponds to the occurrence of a Josephson coupling between the two electrodes, strong enough for the critical current to resist thermal fluctuations. Below $T_{J}$, the critical current $I_c$ grows quadratically when lowering temperature, as expected from Josephson coupling by the proximity effect \cite{degennes}. A third characteristic temperature $T'_c$ is also observed when the barrier itself becomes superconducting. Its existence is inherent to the irradiation fabrication technique which lowers the $T_c$ of the material in the region below the slit.  To retrieve this temperature, we measured the R(T) curve while illuminating the junction with a sufficiently high-power RF signal to suppress the Josephson supercurrent (Fig. 3a). The temperature at which the resistance reaches zero defines $T'_c$. The Josephson regime therefore lies between $T'_c$=45K and  $T_{J}$  = 66 K. \\ 

 \begin{figure}[h]
\includegraphics[width=8.5cm]{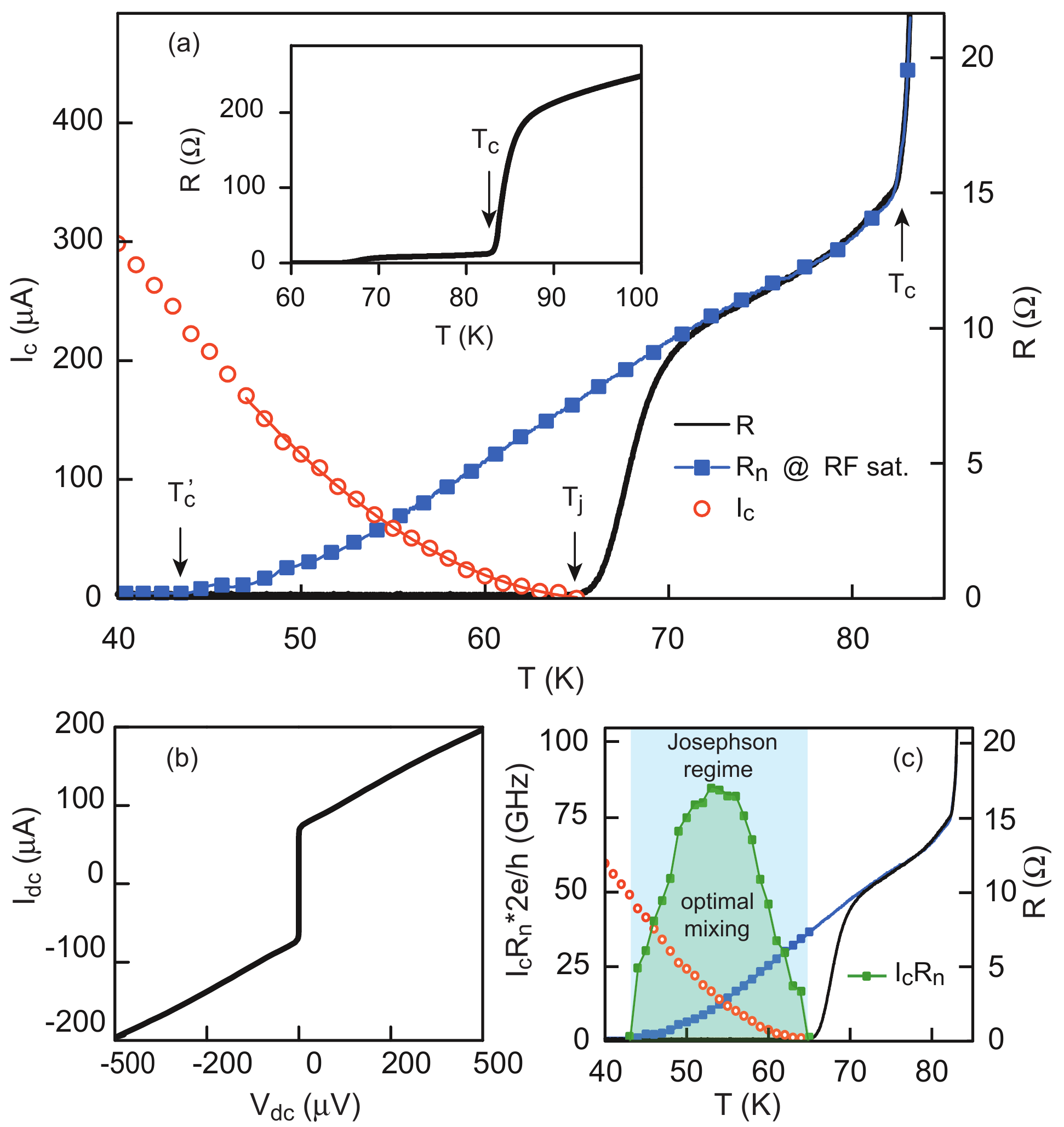}
    \caption{ (a) Resistance $R$, critical current $I_c$ and normal state resistance $R_n$ of the junction as a function of temperature. $R_n$ is obtained by saturating the junction with microwaves. The three temperatures $T_c$, $T_J$ and $T'_c$ are indicated on the graph. Inset : R(T) curve at larger scale. (b)  I(V) curves at T=54 K in the Josephson regime. (c) $I_cR_n$ product in frequency unit superimposed to the previous curves.\\}
 \end{figure}

\indent Junctions have non-hysteretic current-voltage characteristics with an upward curvature  in the dissipation branch at low voltage and no sharp feature at the gap voltage (Fig. 3b).  In this low capacitance regime, the electrical behavior of the junction is expected to be well described by the resistively shunted junction (RSJ) model \cite{mccumber}, which considers a Josephson element in parallel with a resistance $R_n$. The voltage across the junction is given by the  expressions :
\begin{eqnarray}
V&=&R_n\big[I_{\mathrm{dc}}+\delta I_{n}-I_{c}sin\phi\big]\\
\label{RS}
V&=&\frac{\hbar}{2e}\frac{d\phi}{dt}
\label{RSJbis}
\end{eqnarray}

\noindent where $I_{dc}$ is the bias dc current and $\phi$ the superconducting phase difference across the junction. Here $\delta I_{n}$ is an additive Gaussian  white noise of variance $\sigma^2=\frac{\hbar\Gamma}{eI_cR_n\Delta t}$, where $\Gamma=2ek_BT/\hbar I_c$ is the ratio of the thermal energy to the Josephson energy and $\Delta t$ is the time step chosen for the numerical integration of equations (1) and (2). From the determination of $I_c$ and $R_n$, we extract the characteristic frequency $f_c = (2e/h)I_cR_n$ of the mixer.  As seen in Fig. 3c, $f_c$ displays a dome as a function of the temperature with a maximum value $f_c^{\mathrm{opt}}$ of  85 GHz at 55 K. Note that $f_c$ is not a cut-off frequency and that mixing can be performed up to frequencies corresponding to several times the value of $f_c$ at the cost of a reduced conversion efficiency \cite{malnou}. However, for optimal operation, it is desirable to have $f_c$ larger than the frequencies of the incoming signals $f_{\mathrm{LO}}$ and $f_{s}$ as the resulting ac current would then interact mainly with the Josephson non-linear inductive element.\\ 
    \begin{figure}[t]
\includegraphics[width=9cm]{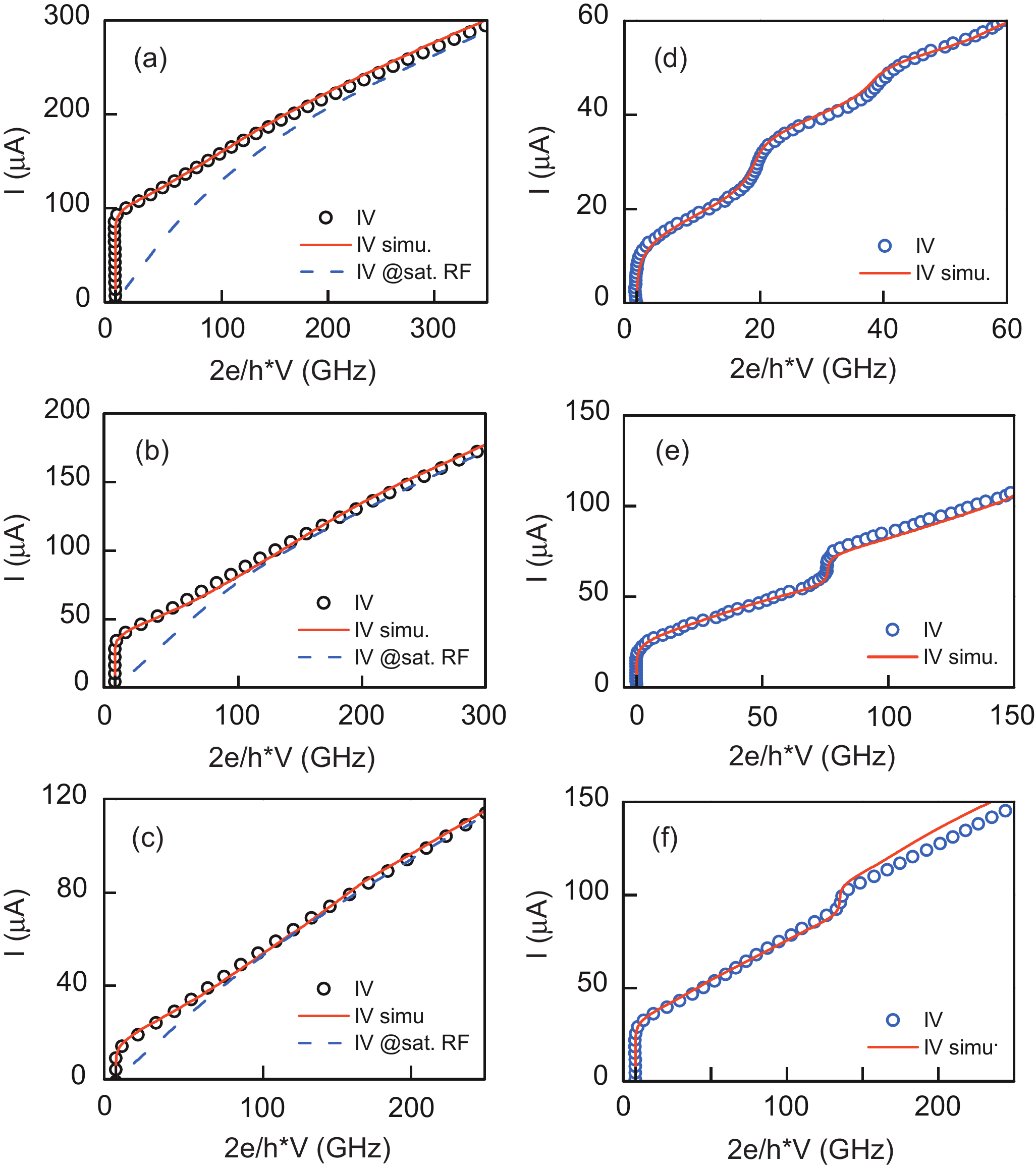}
    \caption{a, b and c) Current-voltage characteristics of the junction (open circles) measured at different temperatures 53K, 58K and 62K respectively. Dashed lines correspond to the curve under strong microwave radiation and red solid lines correspond to a fit with using the RSJ model (\ref{RSJ2}) in which the non-linear resistance is introduced.  d--f) Current voltage characteristics of the junction (open circles) measured at T=58K under LO radiation at 20 GHz, 70 GHz and 140 GHz. Curves are fitted using the RSJ model  (\ref{RSJ2}).}
 \end{figure}
   \begin{figure}[h]
\includegraphics[width=9cm]{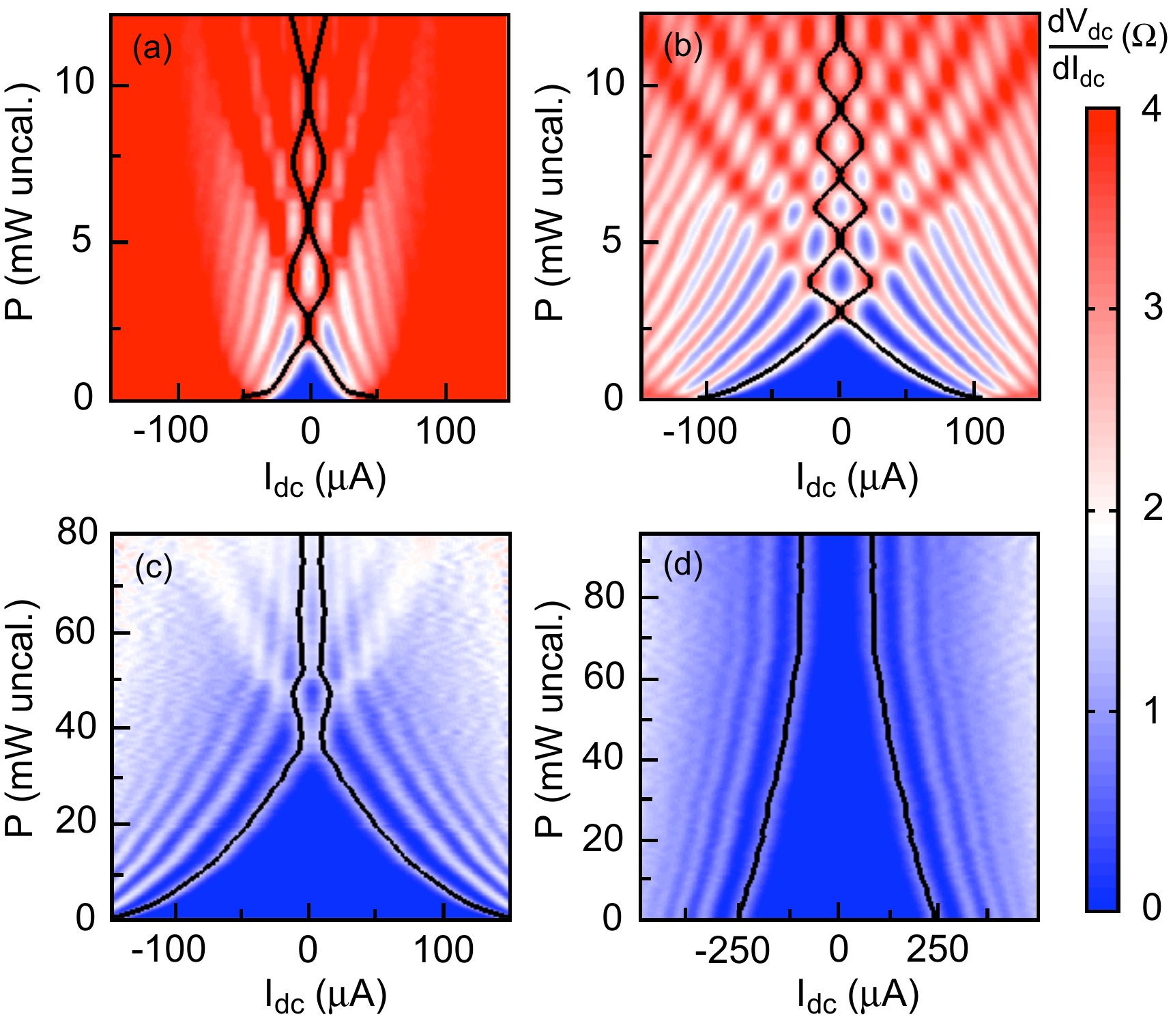}
    \caption{Differential resistance of junction (color scale) under a 20-GHz  LO signal as a function of bias current and RF power for different temperatures : (a) T=56K; (b) T=53K; (c)  T=42K; (d) T=35K. The critical current (n=0 step) as a function of RF power is shown in full  back  line. Complete oscillations of the current height of the Shapiro steps can be seen only in the Josephson regime $T'_c<T<T_J$ (panels (a)--(c)). In the flux flow regime, the critical current is never reduced to zero. }
 \end{figure}

 \indent In the Josephson regime, the dissipation branch at large bias reveals that the resistance increases with voltage. The origin of this non-linear behavior of the resistance stems from the non-uniform distribution of defects in the barrier resulting from the irradiation process \cite{katz}. In Figure 4 we show that the non linear resistance, in particular at low bias, can be retrieved by suppressing the Josephson supercurrent with high-power microwave radiation (dashed lines). The behaviour of the I(V) is then well described by the  RSJ equations (1) and (2) provided we enter the non-linear normal resistance $R_n(I_{dc})$ in the model (Fig. 4). \\

\indent Current-voltage characteristics measured at T= 58K upon LO illumination are shown in Fig. 4(d)--(f) for three different frequencies, 20 GHz, 70 GHz and 140 GHz. Shapiro steps at the quantized voltage $V_n=n\frac{\hbar}{2e}f_{\mathrm{LO}}$ can be clearly observed \cite{shapiro}. To analyze these features, we added a LO current term into equation (1)
\begin{eqnarray}
V=R_n(I_{\mathrm{dc}})\big[I_{\mathrm{dc}}+I_{\mathrm{LO}}\cos(2\pi f_{\mathrm{LO}}t)+\delta I_{n}-I_{c}sin\phi\big]
\label{RSJ2}
\end{eqnarray}
  A good agreement with the experimental data is obtained as can be seen in Fig. 4(d)--(f). The junction response to 20-GHz LO illumination has also been measured at different temperatures.  Fig. 5 shows the differential resistance of the junction $\frac{dV}{dI}$ as a function of bias current and power radiation in the Josephson regime (Fig. 5(a) and (b)) and below (Fig. 5(c) and (d)).   
  For strong LO power, several Shapiro steps can be seen as well as their modulation with LO  power. In particular, the critical current (n=0) can be fully suppressed by the application of the correct amount of LO power. However, below $T'_c$ the modulation of the critical current is no longer complete (Fig. 5(d)), indicating that the dynamics of the junction deviates from a pure Josephson one. A crossover towards a flux flow regime is then observed although some features of the Josephson effect remain observable.\\ 
     \begin{figure}[h]
\includegraphics[width=9cm]{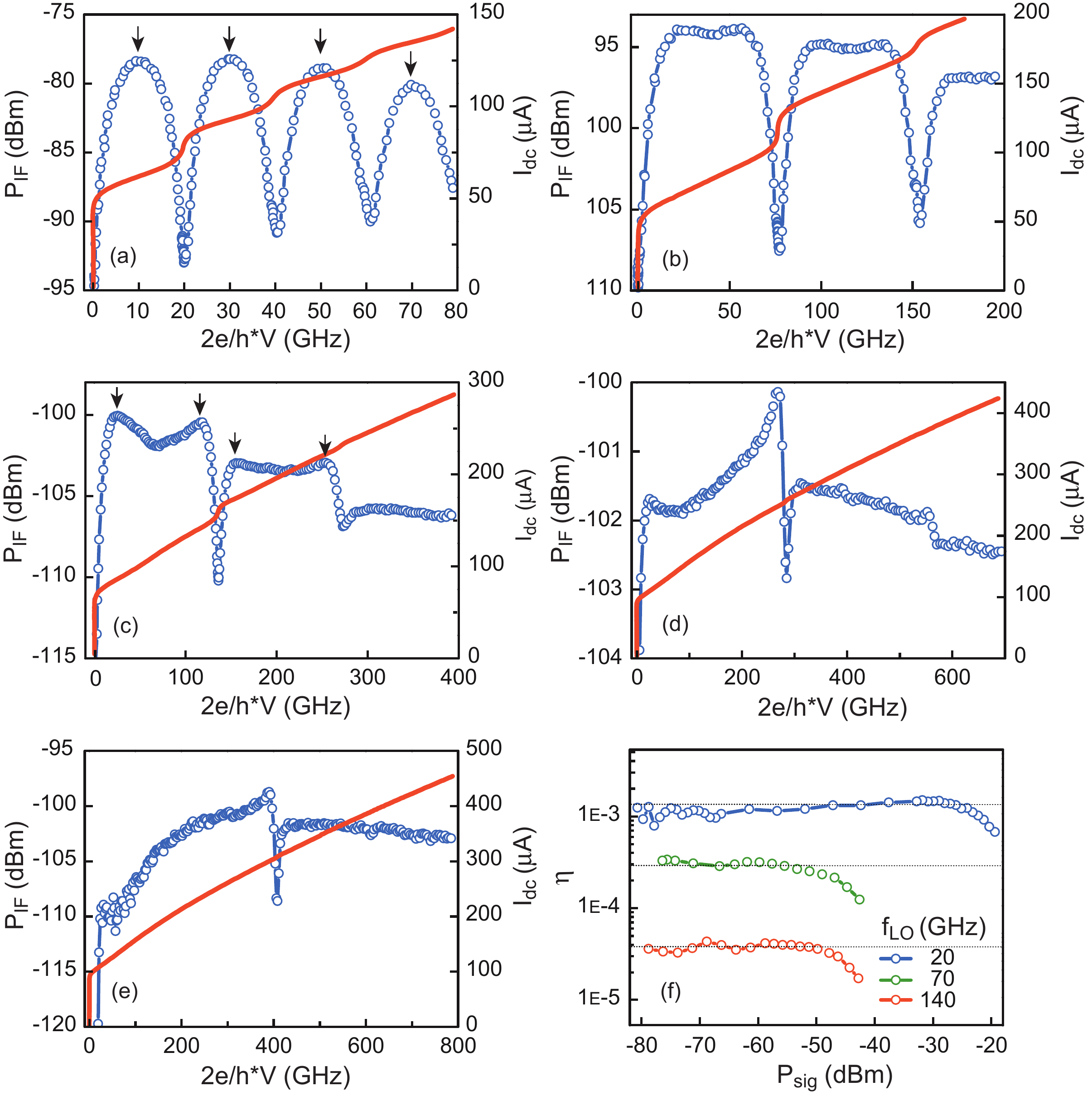}
    \caption{ (a)--(e) Output power at the IF (left scale) and dc current (right scale) as a  function of voltage measured at T=53K for five different LO frequencies, $f_{\mathrm{LO}}$= 20 GHz (a), $f_{\mathrm{LO}}$= 70 GHz (b), $f_{\mathrm{LO}}$= 140 GHz (c), $f_{\mathrm{LO}}$= 280 GHz (d), $f_{\mathrm{LO}}$= 410 GHz (e). The IF frequency is 6 GHz. For the three lowest frequencies (panels (a)--(c), the power of the signal has been set to approximatively one thousandth of the LO power. For the  two highest frequencies (d) and (e), the signal power is of the same order as that for the LO. (f) Conversion efficiency $\eta=\frac{P_{IF}}{P_{\mathrm{s}}}$ measured as a function of the signal power expressed in uncalibrated dBm. The horizontal black lines correspond to the ideal linear response of the mixer. }
 \end{figure}

\textbf{IV. High frequency mixing}\\

   The junction is illuminated with a strong LO signal at frequency $f_{\mathrm{LO}}$ and a much weaker test signal at frequency $f_{\mathrm{s}}$. These conditions guarantee that the IF signal is produced by a first order mixing mechanism between the signal  and the LO. Figure 6 shows the output power measured at the intermediate frequency  $f_{\mathrm{IF}}=|f_{\mathrm{LO}}-f_{s}|$=6 GHz as a function of the dc voltage $V$ across the junction for the different ranges of frequency. At 20 GHz, 70GHz and 140 GHz, the power of the LO has been set to reduce the critical current to approximately half its value, as it corresponds to an optimal operation point for mixer performances (see part VI). The IF output power $P_{\mathrm{IF}}$ displays strong modulations whose period is given by the quantized voltage $\Delta V=\frac{\hbar}{2e} f_{\mathrm{LO}}$ between two Shapiro steps. Two mixing regimes can be identified. For $f_{\mathrm{LO}}$=20GHz (Fig. 6(a)),  $P_{\mathrm{IF}}$  is maximum at voltages corresponding to the exact center between two Shapiro steps (see arrow). We will show in part V that such a behavior is obtained when $f_{\mathrm{LO}}<f_c^{\mathrm{max}}$. For $f_{\mathrm{LO}}$=140GHz (Fig. 6(c)),  $P_{\mathrm{IF}}$  has two maxima close to the Shapiro steps (see arrows), separated by a dip. This corresponds to the condition $f_{\mathrm{LO}}>f_c^{\mathrm{max}}$ . In the intermediate situation where $f_{\mathrm{LO}}\approx f_c^{\mathrm{max}}$, $P_{\mathrm{IF}}$ is approximately flat at the center of the steps (Fig. 6(b)). Measurements performed at higher frequencies, $f_{\mathrm{LO}}$=280 and 410 GHz (Fig. 6(d) and 6(e)) indicate that the junction responds in the lower part of the THz range. However, in these cases the power of the LO source was not sufficient to reach optimal bias conditions. Mixing at frequencies higher than 410 GHz was not investigated in this study.\\
\indent  The output power $P_{\mathrm{IF}}$ at the intermediate frequency  was measured as a function of the signal power for the three main ranges of frequency. After calibration of the IF output line, the conversion efficiency $\eta=\frac{P_{\mathrm{IF}}}{P_{\mathrm{s}}}$ was calculated and plotted as a function of the signal power $P_{\mathrm{s}}$. The mixer displays a linear dynamical range of constant conversion efficiency of more than 55 dB at 20 GHz and 30 dB at 140 GHz (Fig. 6(f)). For strong signal power, the amplitude of the modulation of the IF signal  decreases and  the mixer saturates. In this situation,  the signal power can no longer be considered to be small compared to the LO power and second-order mixing processes take place.\\

        \begin{figure}[tb]
\includegraphics[width=8cm]{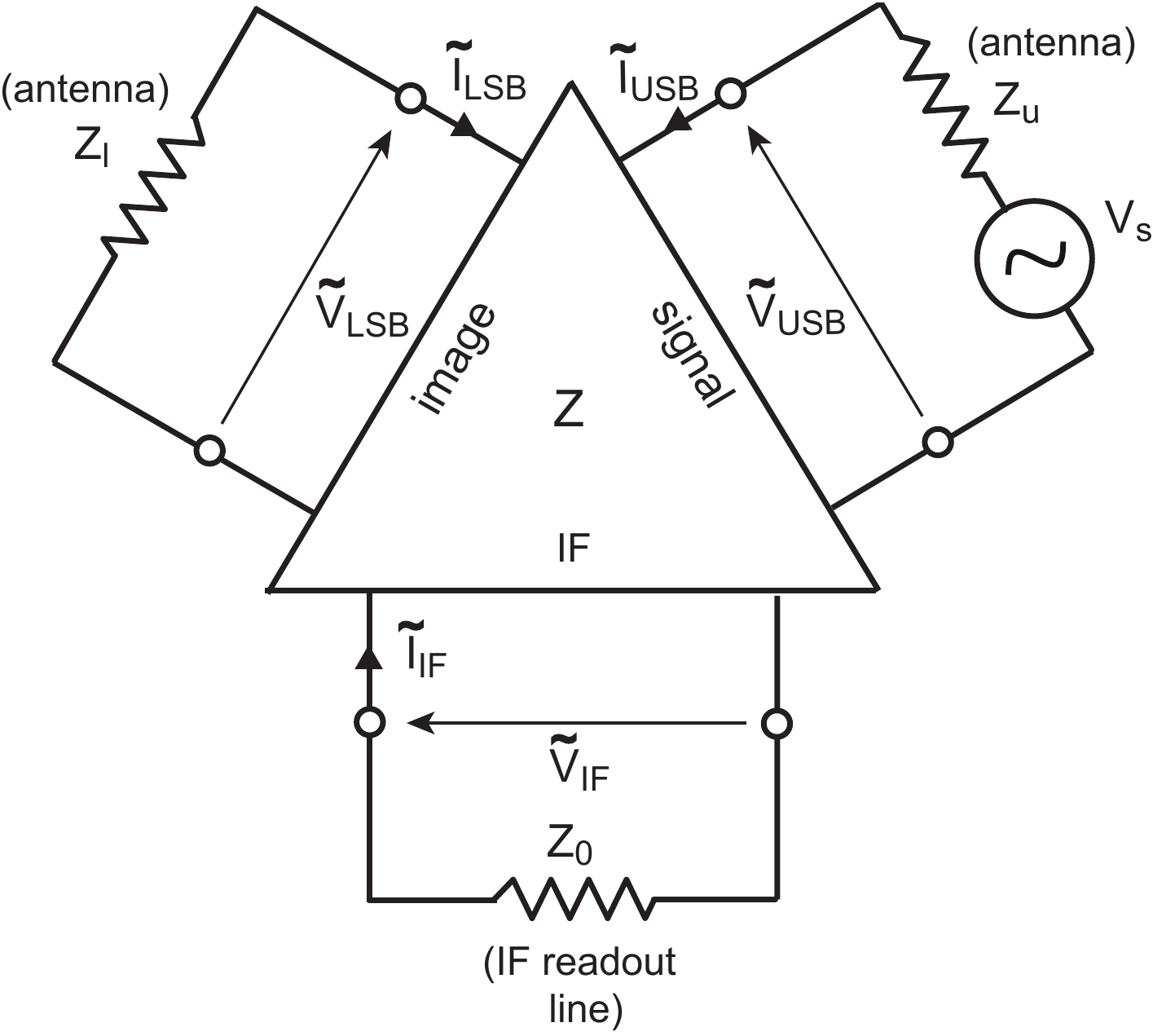}
    \caption{ Equivalent circuit of the Josephson mixer connected to the external impedances $Z_u$, $Z_l$ and $Z_0$.}
 \end{figure}

 \textbf{V. Three-port model\\}
 
Following the pioneering work of Taur \cite{taur}, we used the three-port model to calculate the performance of the mixer. It describes the linear response for a small signal by solving the non-linear response of the mixer under the LO illumination. When  the junction is driven by a strong LO signal, the relation between small currents $I$ and voltages $V$ is linear in the frequency domain \cite{likharev} :
\begin{equation}
\tilde{V}(f)=\sum_{k=-\infty}^{^{+\infty}}Z_{k}(f-kf_\mathrm{LO})\tilde{I}(f-kf_\mathrm{LO})\label{eq:vtild}
\end{equation}

Let us now consider the case when a signal of frequency $f_{\mathrm{s}}$ close to $f_{\mathrm{LO}}$ is shined onto the junction.
At first order in (\ref{eq:vtild}), there are only three frequencies of interest each containing a term at the intermediate frequency  $f_{\mathrm{IF}}=|f_{s}-f_{\mathrm{LO}}|$ : the lower side band frequency  $f_{\mathrm{LSB}}=f_{\mathrm{LO}}-f_{\mathrm{IF}}$, the intermediate frequency itself $f_{\mathrm{IF}}$ and the upper side band one $f_{\mathrm{USB}}=f_{\mathrm{IF}}+f_{\mathrm{LO}}$. 
Limiting ourselves to these three frequencies, (\ref{eq:vtild}) can be written as a matrix equation:
\begin{eqnarray}
\left(\begin{array}{c}
\tilde{V}_{\mathrm{USB}}\\
\tilde{V}_{\mathrm{IF}}\\
\tilde{V}_{\mathrm{LSB}}^{*}
\end{array}\right)=\left[\begin{array}{ccc}
Z_{uu} & Z_{u0} & Z_{ul}\\
Z_{0u} & Z_{00} & Z_{0l}\\
Z_{lu} & Z_{l0} & Z_{ll}
\end{array}\right]\left(\begin{array}{c}
\tilde{I}_{\mathrm{USB}}\\
\tilde{I}_{\mathrm{IF}}\\
\tilde{I}^{*}_{\mathrm{LSB}}
\end{array}\right)\label{eq:3ports}
\end{eqnarray}

 \noindent where $u$, $l$, and $0$ stand for USB, LSB, and IF respectively. $\tilde{\tilde{Z}}$ is the impedance matrix of the mixer which characterizes in particular its ability to down-convert at $f_{IF}$ the information at $f_{\mathrm{USB}}$ ( or $f_{\mathrm{LSB}}$. Each of its elements $Z_{ij}$ is simply the ratio of the voltage $\tilde{V}_j $ at  frequency $f_j$ to the current $\tilde{I}_i$
injected at frequency $f_i$. For a low-Q mixer, the symmetric properties of the 3-port matrix imply $Z_{uu}=Z_{ll}^{*}$, $Z_{lu}=Z_{ul}^{*}$,
$Z_{u0}=Z_{l0}^{*}$ and $Z_{0u}=Z_{0l}$.  A general mixer theory  provides the following expression for the matrix elements \cite{torrey,taur, schoelkopf}

\small
\begin{eqnarray}
 Z_{uu}&=&\frac{1}{2}\Big[\frac{\partial\tilde{V}(f_{\mathrm{LO}})}{\partial I_{\mathrm{LO}}}+\frac{\tilde{V}(f_{\mathrm{LO}})}{I_{\mathrm{LO}}}\Big]\,(\mathrm{RF}\,\mathrm{impedance})\\
Z_{u0}&=&\frac{\partial\tilde{V}(f_{\mathrm{LO}})}{\partial I_{dc}}\,(\mathrm{up}\,\mathrm{conversion})\\
Z_{ul}&=&\frac{1}{2}\Big[\frac{\partial\tilde{V}(f_{\mathrm{LO}})}{\partial I_{\mathrm{LO}}}-\frac{\tilde{V}(f_{\mathrm{LO}})}{I_{\mathrm{LO}}}\Big]\,(\mathrm{image}\,\mathrm{conversion})\\
Z_{0u}&=&\frac{1}{2}\frac{\partial V_{dc}}{\partial I_{\mathrm{LO}}}\,(\mathrm{down}\,\mathrm{conversion})\\
Z_{00}&=&\frac{\partial V_{dc}}{\partial I_{dc}}\,(\mathrm{dc}\,\mathrm{impedance})
\end{eqnarray}
\normalsize

To determine the conversion efficiency of the mixer, we introduce in Fig. 7 the external part of the circuit which is described by  the diagonal impedance matrix $\tilde{\tilde{Z}}_{\mathrm{ext}}$ whose  elements $Z_u$, $Z_l$ and $Z_0$ are connected to the mixer inputs. Here $Z_u$ and $Z_l$ represent the impedance of the spiral antenna (80 $\Omega$) at USB and LSB frequencies respectively and are taken to be identical.  $Z_0$ is the  50-$\Omega$ impedance of the IF microwave readout line.  Assuming that the signal $V_{\mathrm{s}}$ incoming on the antenna is at the USB frequency, the equation for the circuit shown in Fig. 7 is \cite{taur}
\begin{eqnarray}
\left(\begin{array}{c}
\tilde{V}_{\mathrm{USB}}\\
\tilde{V}_{\mathrm{IF}}\\
\tilde{V}_{\mathrm{LSB}}^{*}
\end{array}\right)+\left[\begin{array}{ccc}
Z_{u} & 0 & 0\\
0 & Z_{0} & 0\\
0 & 0 & Z_{l}
\end{array}\right]\left(\begin{array}{c}
\tilde{I}_{\mathrm{USB}}\\
\tilde{I}_{\mathrm{IF}}\\
\tilde{I}^{*}_{\mathrm{LSB}}
\end{array}\right)=\left(\begin{array}{c}
V_{\mathrm{s}}\\
0\\
0
\end{array}\right)
\label{ext}
\end{eqnarray}

We therefore obtain a relation between the currents at different frequencies and the input signal 

\begin{eqnarray}
\left(\begin{array}{c}
\tilde{I}_{\mathrm{USB}}\\
\tilde{I}_{\mathrm{IF}}\\
\tilde{I}^{*}_{\mathrm{LSB}}
\end{array}\right)=\tilde{\tilde{Y}}
\left(\begin{array}{c}
V_{\mathrm{s}}\\
0\\
0
\end{array}\right)
\label{tot}
\end{eqnarray}

where $\tilde{\tilde{Y}}=(\tilde{\tilde{Z}}+\tilde{\tilde{Z}}_{\mathrm{ext}})^{-1}$ is the admittance matrix.

 We define the conversion efficiency as the ratio of the IF power $P_{IF}=\frac{1}{2}Z_0|\tilde{I}_{\mathrm{IF}}|^2$ dissipated in the impedance $Z_0$ to the available signal power $P_{s}=\frac{\left|V_{\mathrm{s}}\right|^{2}}{8Z_{u}}$ on the antenna impedance $Z_u$,
 \begin{eqnarray}
\eta_{c}=\frac{P_{IF}}{P_{\mathrm{s}}}=4Z_0Z_{u}\left|Y_{u0}\right|^{2}
   \label{conversion}
   \end{eqnarray}
   
    where $Y_{u}^{0}$ is a non-diagonal matrix element of the admitance matrix $\tilde{\tilde{Y}}$ with  the same convention as in (\ref{eq:3ports})
 
 In the limit---well satisfied experimentally--- where $|Z_{ul}|,|Z_{u0}|\ll|Z_{u}|$ and $|Z_{0u}|\ll|Z_{0}|$, the conversion efficiency takes the simple form
 
  \begin{eqnarray}
\eta_{c}=4\frac{Z_{u}}{|Z_{u}+Z_{uu}|^2}\times\frac{Z_{0}}{|Z_{0}+Z_{00}|^2}\times Z_{0u}^2
   \label{conv_simple}
   \end{eqnarray}
 
The first two factors correspond to the matching impedance conditions at the USB and IF frequencies. The conversion is optimal when the antenna impedance $Z_{u}$ matches the RF impedance of the junction $Z_{uu}$ and when the readout line impedance $Z_{0}$ matches the dc impedance of the junction.
The last factor $Z_{0u}^2$ represents the ability of the junction to down convert the USB signal to the intermediate frequency. Within this approximation,  the performance of the device mainly depends on three elements of the $\tilde{\tilde{Z}}$ matrix : (i) the RF impedance at the USB (or LSB) frequency $Z_{uu}$ (=$Z_{ll}^{*}$), (ii) the dc impedance  $Z_{00}$, and (iii) the down-conversion impedance  $Z_{0u}$ (=$Z_{0l}^{*}$). \\
       \begin{figure}[b]
\includegraphics[width=9cm]{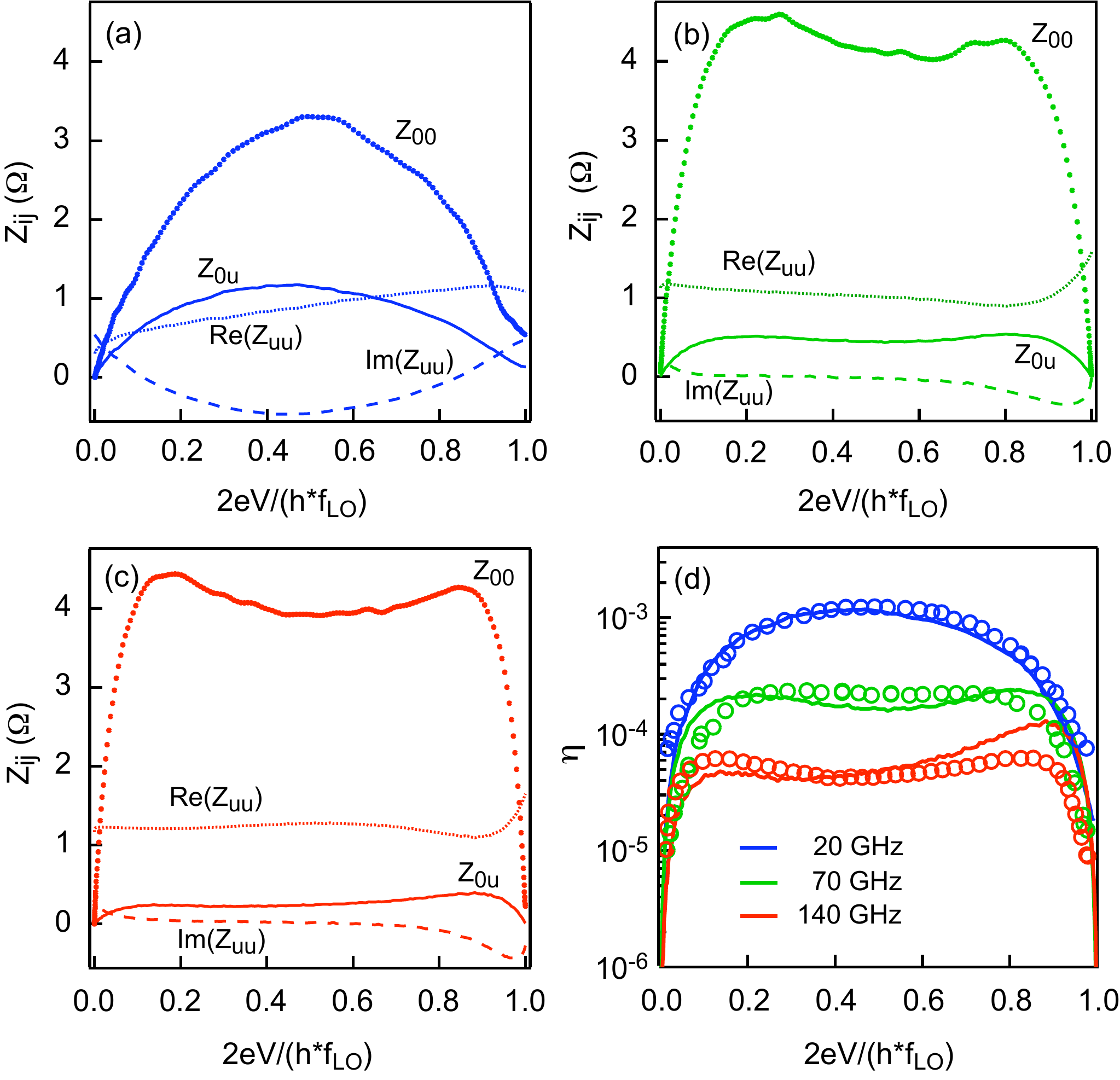}
    \caption{Main elements of the scattering matrix as a function of normalized voltage calculated at T=58K for $f_\mathrm{LO}$=20 GHz (panel a), $f_\mathrm{LO}$=70 GHz (panel b) and $f_\mathrm{LO}$=140 GHz (panel c). d) Comparison between experimental (dots) and theoretical (full lines) conversion efficiency $\eta_C$ calculated for the three LO frequencies. }
 \end{figure}

To derive the impedance matrix $\tilde{\tilde{Z}}$,  the RSJ equation (\ref{RSJ2}), including the non-linear resistance $R_n(I_{dc})$ is first solved numerically in the time domain. The biasing condition in terms of LO power is chosen to reproduce the experimental conditions of Fig. 6. The matrix elements $Z_{ij}$ are then calculated according to expressions (6), (9) and (10)  and plotted in Fig. 8 as a function of normalized voltage for the different LO frequencies.
The impedances $Z_{00}$  and $Z_{0u}$ reproduce the shape of the  output power $P_{\mathrm{IF}}$  of Fig. 6. For $f_{LO}$=20GHz  the mixer should be dc biased halfway between the Shapiro steps whereas for $f_{\mathrm{LO}}$=140 GHz it should be biased close to the steps.  The impedance  $Z_{0u}$ and therefore the ability of the junction to down-convert decreases significantly when the LO frequency is increased. Figure 8(d) shows that the  theoretical calculations of the conversion efficiency obtained from (\ref{conv_simple}) are in good agreement with experimental data.  A crossover from the first regime of mixing $f_\mathrm{LO}<f_c$   to the second regime $f_\mathrm{LO}>f_c$ is observed.  At T=53K, the noise parameter $\Gamma$=0.07 is much lower than 1 which guarantees  that the Josephson non-linearity is not smeared out by the noise.\\

 \indent The conversion efficiency takes a maximum value of 0.1\% at 20GHz and decreases to 0.01\% at 140GHz. An improvement of the mixer performances  requires optimizing the three factors of expression (\ref{conv_simple}).  In particular, the impedance mismatch resulting from the low values of $Z_{uu}$ and $Z_{00} $ compared with $Z_u$ and $Z_0$  respectively, leads to a significant deterioration of $\eta_c$.  In practice, the matrix elements are determined by two parameters, the normal resistance $R_n$ of the junction and its characteristic frequency $f_c$ (i.e. the $I_cR_n$ product), through the RSJ equation. As $R_n$ is the only impedance entering this equation, all the matrix elements are directly proportional to it. This resistance  needs to be increased significantly  to improve the impedance matching.   This can be done by decreasing both the width and the thickness of the junction and by increasing the ion irradiation fluence.  Finally, impedance matching elements both between the antenna and the junction and between the readout line and the junction could also be added at a cost of reduced bandwidth.\\
 
 The value of $f_c$ influences all the matrix elements, but affects mainly the down-conversion one $Z_{0u}$. Assuming for simplification that $Z_{uu}\sim Z_{00}\sim R_n$, the amount of LO current necessary to reduce the critical current to zero is $\Delta I_{LO}\approx\frac{hf_\mathrm{LO}}{2eR_n}$ \cite{grimes}. We thus obtain the dependence of the $Z_{0u}$ element with the ratio $f_c$/$f_\mathrm{LO}$
  \begin{equation}
Z_{0u}=\frac{1}{2}\frac{\partial V_{dc}}{\partial I_{\mathrm{LO}}}\sim\frac{1}{2}\frac{\Delta (R_nI_c)}{\Delta I_\mathrm{LO}}\sim \frac{1}{2}R_n\frac{f_c}{f_\mathrm{LO}}
\end{equation}
 From this expression, we see that it is desirable to fabricate junctions with high  $f_c$ values, i.e.  high $I_cR_n$ product.
 The junction presented in this study has a characteristic frequency which is lower than the ones usually reported in grain-boundary junctions\cite{chen,scherbel,harnack,tarasov}. However, several developments can be made to optimize the $I_cR_n$ product \cite{sirena, sirenaJAP2007} in our junctions. In particular, a higher irradiation fluence combined with an annealing of the sample should lead to a significant improvement  \cite{sirenaAPL2007,sirenaJAP2009}.  For $f_\mathrm{{LO}}>>f_{c}$  the signal and the LO ac current interact weakly with the inductive Josephson element. As a result, a large part of the  IF power is generated by mixing on the non-linear resistance. As can be seen in Fig.  6(d) and (e), this produces a continuous background on top of which, Josephson mixing can still be distinguished. \\
     \begin{figure}[t]
\includegraphics[width=9cm]{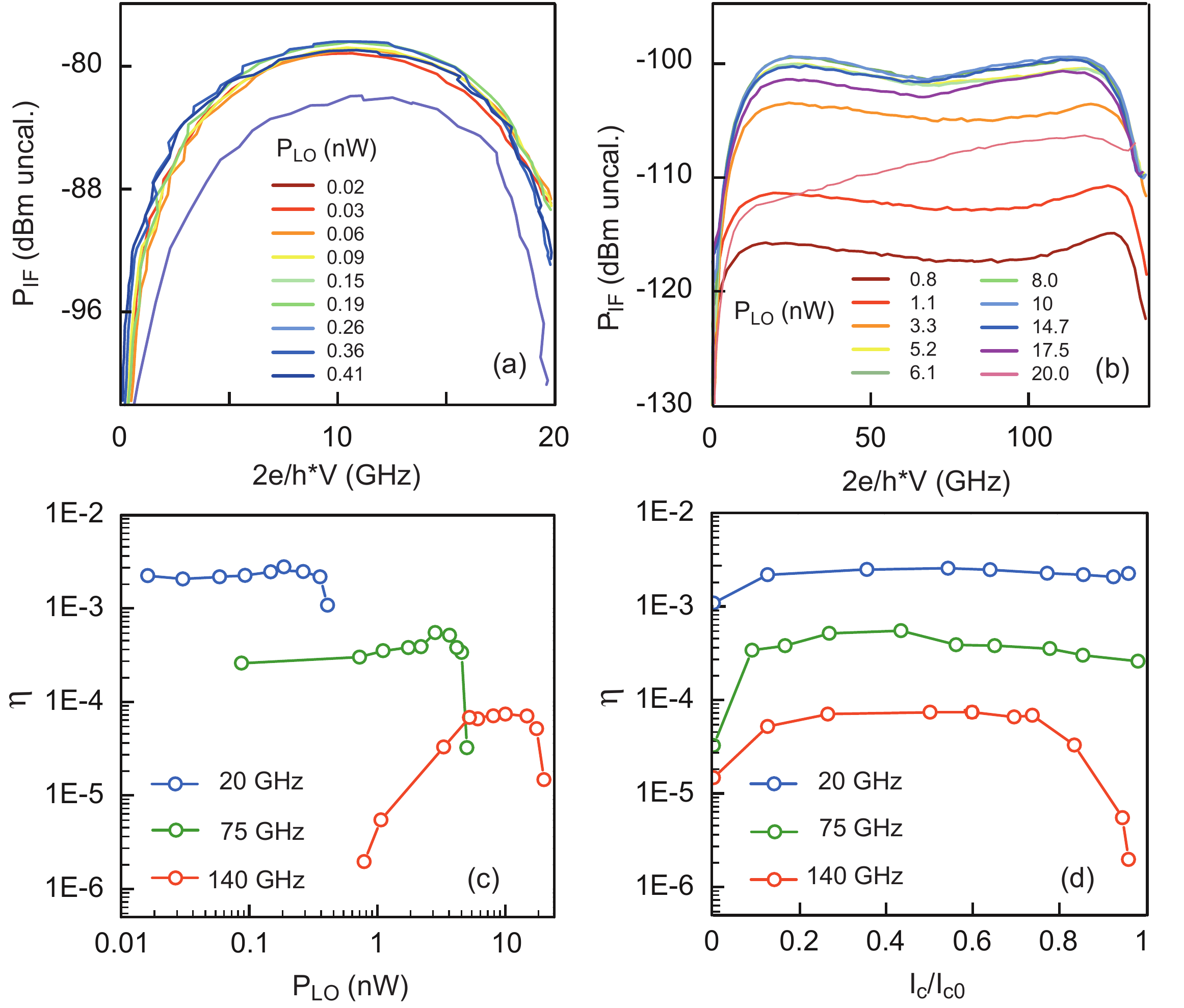}
    \caption{(a) and (b) Power in dBm at IF measured for different values of the LO power received by the junction.  The LO frequency is 20 GHz on panel (a) and 140 GHz on panel (b). (c)  Conversion efficiency taken at $(2e/h)V=f_\mathrm{{LO}}/2$ (circle) and  $2e/h\times V=3f\mathrm{{LO}}/2$ (square) as a function of the LO power coupled to the junction. d) Conversion efficiency taken at $2e/h\times V=f_\mathrm{{LO}}/2$ (circle) as a function of critical current reduction  for the three LO frequencies. }
 \end{figure}

  \textbf{VI. Influence of the LO power\\}
  
In a practical heterodyne receiver application, the LO power necessary to optimally bias the mixer is a critical parameter and must satisfy two important requirements :
 (i) it has to be as low as possible to minimize the power consumption and to be easily driven by available sources in the frequency range of interest and (ii) its variations and fluctuations must not modify significantly the performance of the mixer. For a Josephson mixer, the dependence of the conversion efficiency with the LO power is mainly determined by the characteristic frequency. Additionally, it is generally expected that the conversion should be greatest for a LO power corresponding to a suppression by approximately 50$\%$ of the critical current.  However, a careful analysis of this point has never been done, and the mixer should in principle operate for a range of LO power. Figure 8(a) and (b) show the behavior of the output power $P_\mathrm{{IF}}$ as a function of voltage across the junction for different values of the LO power received by the junction, for $f_\mathrm{{LO}}$= 20 GHz and 140 GHz. The conversion efficiency taken at $2e/h\times V=f_\mathrm{{LO}}/2$  is plotted as a function of $P_\mathrm{{LO}}$ (Fig. 8(c)). For $f_\mathrm{{LO}}<f_c$, $\eta$ is constant on more than one decade and decreases at  strong LO power. $P_\mathrm{{LO}}$ as low as 20 pW at $f_{LO}$= 20 GHz  and 100 pW at $f_{LO}$= 70 GHz are sufficient to drive optimally the mixer whereas at 140 GHz, 10 nW of power are required. It is clear that the conversion efficiency does not depend critically on the LO power as long as $f_\mathrm{{LO}}< f_c$; otherwise,  as can been seen at 140 GHz, $\eta$ is optimal for a given LO power  which corresponds approximately to a suppression by 50$\%$ of the critical current (Fig. 8(d)).

\indent In conclusion, we have demonstrated the mixing operation of ion-irradiated \YBCO Josephson junctions up to 420 GHz at temperature higher than 50K. The performances of the mixer were studied as a function of  LO frequency and LO power. For LO frequencies lower or close to the characteristic frequency of the junction, a conversion efficiency in the range of  0.02-0.1\%  was obtained for a LO power lower than 1 nW. A detailed analysis of the mixer within the framework of the general three-port model  and the RSJ model  was proposed. A good agreement between experimental data and numerical simulation was obtained.\\

\indent The authors thank T. Ditchi, E. Geron, S. Hole and J. Lucas for useful discussions, and Y. Legall for his help with the ion irradiations. This work was supported by the ANR ASTRID program, the  Emergence program Contract of Ville de Paris and by the R\'egion Ile-de-France in the framework of CNano IdF and Sesame program. CNano IdF is the nanoscience competence center of the Paris Region, supported by CNRS, CEA, MESR and R\'egion Ile-de-France. \\

 \end{document}